\documentclass[12pt]{article}

\usepackage{cite}
\usepackage{graphicx}

\def\beq{\begin{equation}}
\def\eeq#1{\label{#1}\end{equation}}
\def\eeqn{\end{equation}}
\def\beqa{\begin{eqnarray}}
\def\eeqa#1{\label{#1}\end{eqnarray}}
\def\eeqan{\end{eqnarray}}
\def\CR{\nonumber \\ }
\def\leqn#1{(\ref{#1})}

\def\eps{\epsilon}
\def\xsq{\left<|X|^2\right>}
\def\goesto{\rightarrow}
\def\stacksymbols #1#2#3#4{\def\theguybelow{#2}
    \def\vp{\lower#3pt}
    \def\sp{\baselineskip0pt\lineskip#4pt}
    \mathrel{\mathpalette\intermediary#1}}

\def\intermediary#1#2{\vp\vbox{\sp
     \everycr={}\tabskip0pt
     \halign{$\mathsurround0pt#1\hfil##\hfil$\crcr#2\crcr
              \theguybelow\crcr}}}

\def\gapproxeq{\stacksymbols{>}{\sim}{2.5}{.2}}
\def\lapproxeq{\stacksymbols{<}{\sim}{2.5}{.2}}

\begin{document}
%%%%%%%%%%%%%%%%%%%%%%%%%%%%%%%%%%%%%%%%%%%%%%%%%%%%%%%%%%%%%%%%%%%%%%%%%%%%

\begin{titlepage}
%%%%%%%%%%%%%%%%%%%%%%%%%%%%%%%%%%%%%%%%%%%%%%%%%%%%%%%%%%%%%%%%%%%%%%%%%%%%%%%
\begin{flushright}
UCB-PTH-02/53 \\
LBNL-51663 \\
{\tt hep-ph/0211369} \\
\end{flushright}

\vskip.5cm
\begin{center}
{\huge{\bf Preheating \\ in Supersymmetric Theories}}
\vskip.4cm
\end{center}
\vskip0.2cm

\begin{center}
{\sc Z. Chacko}$^{a,b}$,
{\sc H. Murayama}$^{a,b}$
and {\sc M. Perelstein}$^{a}$

\end{center}
\vskip 10pt

\begin{center} $^{a}$ {\it Theory Group, Lawrence Berkeley National 
Laboratory, Berkeley, CA 94720, USA } \\ 
\vspace*{0.1cm}
$^{b}$ {\it Department of Physics, University of California, Berkeley, CA 
94720, USA } \\
\vspace*{0.1cm}  
{\tt zchacko@thsrv.lbl.gov, murayama@hitoshi.berkeley.edu, 
meperelstein@lbl.gov}
\end{center}

\vglue 0.3truecm

\begin{abstract} 
\vskip 3pt 
\noindent
We examine the particle production via preheating at the end of  
inflation in supersymmetric theories. The inflaton and matter scalars are 
now necessarily complex fields, and their relevant interactions are 
restricted by holomorphy. In general this leads to major changes both in the 
inflaton dynamics and in the efficiency of the preheating process. In
addition, supersymmetric models generically contain multiple isolated vacua, 
raising the possibility of non-thermal production of dangerous topological 
defects. Because of these effects, the success of leptogenesis or WIMPZILLA 
production via preheating depends much more sensitively on the detailed 
parameters in the inflaton sector than previously thought.
\end{abstract}

\end{titlepage}
\newpage

%%%%%%%%%%%%%%%%%%%%%%%%%%%%%%%%%%%%%%%%%%%%%%%%%%%%%%
%\setcounter{equation}{0}
\setcounter{footnote}{0}
%%%%%%%%%

\section{Introduction.}

The inflationary paradigm~\cite{book} has been remarkably successful in 
explaining the observed large-scale features of the universe. Apart from
its prediction of a flat universe with $\Omega_{tot}=1$, inflation
naturally produces primordial density fluctuations with a Harrison-Zeldovich 
power spectrum. These fluctuations in turn give rise to the angular pattern of 
the cosmic microwave background radiation (CMBR) consistent with recent 
observations~\cite{data}. 

On the theoretical front, one of the most important recent developments in  
inflationary cosmology was the realization~\cite{Branden, Linde, Linde1} that 
non-perturbative particle production may play an important role in the 
inflaton decay process. The process of the inflaton decay through 
non-perturbative, parametrically enhanced particle production has been 
called {\it preheating}, to distinguish it from the usual reheating scenario
where the inflaton decay is treated perturbatively. An interesting feature
of the preheating scenario is that the production of particles with masses
greater than the inflaton mass is possible. For example, it was claimed
that in the simplest 
chaotic inflation model with the inflaton mass $m \sim 10^{13}$ GeV, the 
scalars with masses of up to $10^{14}-10^{15}$ GeV\footnote{Heavier scalar
particles could be produced via a slightly different mechanism, ``instant''
preheating~\cite{instant}.} and fermions with masses of
up to $10^{17}-10^{18}$ GeV can be copiously produced. In particular, the
non-thermally produced particles could include the baryon number violating 
gauge and Higgs bosons of grand unified theories (GUTs) or 
right-handed neutrinos, resurrecting the possibility of GUT-scale 
baryogenesis or leptogenesis~\cite{KLR, Giudice}. 
%(In supersymmetric models, the bound on the gravitino abundance requires the 
%reheat temperature to be less than about $10^{10}$ GeV, so the GUT-scale 
%particles cannot be produced thermally.) 
The non-thermally produced superheavy
particles could also play the role of dark matter~\cite{WIMPZILLAS!}, and 
their decays could explain the observed super-GZK cosmic ray 
events~\cite{Gelmini}. 

The magnitude of the observed CMBR anisotropies suggests that the energy scale 
at which inflation takes place is substantially lower than the Planck scale. 
Any microscopic model of inflation should explain this hierarchy of scales 
and its radiative stability. From theoretical point of view, the most 
well-established and appealing way to do this is to assume that the physics 
of inflation is supersymmetric. This assumption is independently motivated by 
the need to resolve the usual gauge hierarchy problem of the Standard Model, 
and by the consistency considerations in string theory. Supersymmetry 
puts non-trivial constraints on the spectrum of the theory: bosons and 
fermions necessarily come in pairs, and matching the bosonic and fermionic
degrees of freedom is only possible if all the scalar fields are {\it 
complex}. Moreover, the interaction terms possible in a supersymmetric 
theory are restricted by holomorphy. However, most of the studies of 
the preheating process to date concentrate on the case of a real scalar 
inflaton field, do not take into account the holomorphy constraints on its
couplings, and consider production of fermions and bosons independently.
In this paper, we would like to examine the inflaton dynamics and the 
process of preheating after the end of inflation, taking into account the  
constraints of supersymmetry. We will show that many interesting and 
important new features arise in supersymmetric models, which can substantially 
modify the conventional preheating scenarios. 

Let us mention that the parametric resonance for the case of a complex 
inflaton coupled to a scalar field was considered in Ref.~\cite{complex}. 
However, the structure of the scalar potential assumed in that study is
quite different from the one studied here, which we believe is generic in 
supersymmetric theories. Also, we discuss preheating of fermions which was 
not considered in~\cite{complex}.
  
The paper is organized as follows. We begin by reviewing the well-known 
qualitative results for preheating of scalar particles and fermions in 
the case of a real inflaton (which we refer to as ``non-supersymmetric'') 
in Section~\ref{nonSUSY}. We then proceed to discuss the general 
structure of the inflaton interactions in supersymmetric models, and
explain the assumptions of our analysis (see Section~\ref{IInt}.) In
Sections~\ref{SUSYsc} and~\ref{SUSYf}, we discuss the preheating of scalar 
and fermion particles, respectively, in the supersymmetric case. We  
highlight the similarities and differences between this case and the
more familiar, non-supersymmetric case. We conclude in Section~\ref{Conc}.
Certain more technical points of our analysis are relegated to the 
Appendices. 

\section{Review of Non-Supersymmetric Results.}
\label{nonSUSY}

Before discussing the supersymmetric case, let us recall the 
standard theory of particle production via preheating
in non-supersymmetric models. To be concrete, we consider 
the case of chaotic inflation with the inflaton potential $V(\phi) =
m^2 \phi^2/2$, where the inflaton mass $m = 10^{-6} M_p \approx 10^{13}$ GeV 
and the value of the inflaton immediately after the end of slow-roll 
inflation is $\phi_0 \approx M_p/3$. We will review the non-thermal 
production of scalars and fermions in turn.  

\subsection{Production of Scalar Particles.}
\label{nonSUSY1}

The production of scalar particles during preheating has been studied in
detail in~\cite{Linde1}. Let us briefly summarize the results.
Assume that the inflaton is coupled to a real scalar field $\chi$ via 
\beq
{\cal L}_s \,=\,{1 \over 2}\,(M^2_0 + g^2 \phi^2) \, \chi^2 \,
+V(\phi).
\eeq{scalar}
In a flat FRW universe with a scale factor $a(t)$, the equation for the mode 
of the field $\chi$ with a comoving momentum ${\bf k}$ reads,
\beq 
\ddot{\chi}_k + 3 H \dot{\chi}_k + \omega_k^2 \chi_k = 0,
\eeq{mode}
where 
\beq
\omega_k^2 = {{\bf k}^2\over a(t)^2} + M^2_0 +g^2 \phi^2(t)
\eeq{omega}
is the time-dependent effective frequency of the mode and $H=\dot{a}/a$. It 
is useful to 
first consider particle creation in Minkowsky space, with $a \equiv 1$
and $\phi = \phi_0 \cos mt$. The qualitative features of the process 
depend crucially on the parameter $q \equiv g^2 \phi_0^2 /4m^2$. Typically  
$q \gg 1$ during the preheating epoch. In this case, the $\chi$ particles are
created via broad parametric resonance. In this regime, the necessary 
condition for particle production is non-adiabaticity in the change of the 
effective frequency,
\beq
{d\omega \over dt} \gapproxeq \omega^2. 
\eeq{condition}
This condition can be satisfied if 
\beq
M_0^2 + k^2 \lapproxeq {m g \phi_0 \over 2}.
\eeq{cond1}
That is, particles with bare masses of up to 
\beq
M_{max} \sim \sqrt{mg\phi_0/2} = m q^{1/4}
\eeq{massbound}
are produced. For example, for $g\sim 1$, we find $M_{max} \sim 1.5 \times
10^{15}$ GeV, more than two orders of magnitude above the inflaton mass.  
The maximum momentum for a particle of bare mass $M_0$ is of order $k_{max} 
\sim \sqrt{mg\phi_0/2 - M_0^2}$. Note that for $M$ close to $M_{max}$, only
particles with very low momenta can be produced, so that the production is
suppressed; however, once $M \lapproxeq M_{max}/2$, the typical
momenta of the produced particles become of order $\sqrt{mg\phi_0/2}$, and
the number density of the produced particles becomes of the same order as in 
the case $M_0 = 0$. 

The condition~\leqn{condition} is satisfied and
the $\chi$ particles are produced only for a short period of time during   
each oscillation, when the inflaton is close to the minimum of its 
potential:
\beq
|\phi| \lapproxeq \sqrt{m\phi_0 \over g} = {\phi_0 \over \sqrt{2}} q^{-1/4}
\ll \phi_0. 
\eeq{interval}
For the rest of the oscillation period, the number of $\chi$ particles 
remains constant. Because of this feature, the process of particle   
creation can be treated analytically~\cite{Linde1}.

The expansion of the universe brings in two interesting new features. 
First, the number of particles with a given wavelength may both increase 
and decrease during a particular passage of the inflaton through its minimum, 
depending on the phase of $\chi_k$ at that time. Nevertheless, the number of 
$\chi$ particles is growing exponentially with time when averaged over 
many oscillation periods (this phenomenon has been called ``stochastic''
resonance~\cite{Linde1}) so preheating does occur. Secondly, the amplitude
of the inflaton oscillations is damped with time due to Hubble friction,
so that effectively the $q$ parameter decreases with time. Particle production 
ceases when the condition~\leqn{cond1} is no longer satisfied for any $k$, 
that is, when $q = q_{min} \sim (M_0/m)^4.$ (If $M_0 \lapproxeq m$, 
the preheating stops when $q \sim 1$.) As we have mentioned above, for 
$q \gapproxeq (2M_0/m)^4 =16q_{min}$ the rate of particle production is of 
the same order as in the case $M_0=0$; for $q_{min} \lapproxeq q 
\lapproxeq 16 q_{min}$ there is some production, but the rate is suppressed. 

In the above discussion, we have neglected the backreaction of the 
produced $\chi$ particles on the inflaton motion. If $M_0 = 0$,   
this approximation breaks down when~\cite{Linde1} 
\beq
q \lapproxeq
q_2 = {2 g \mu M_p \over 3m} \ln^{-1} {10^{12} m \over g^5 M_p}. 
\eeq{q2}
where $\mu$ is a parameter of order 0.1 which characterizes the efficiency
of particle production. For most reasonable parameter values, the 
backreaction becomes important after ${\cal O}(10)$ inflaton 
oscillations~\cite{Linde1}. The same estimate is true for massive particles,
provided that their production is unsuppressed for all $q>q_2$. This
occurs if $q_2 \gapproxeq 16 q_{min}$, or, assuming $g\sim 1$, 
\beq
M_0 < {m \over 2} q_2^{1/4} \approx 4m.
\eeq{bkimp}
If, on the other hand, $q_2 < q_{min}$, the backreaction can always be 
neglected: this is the case for $M_0 \gapproxeq 8m = 8\times10^{13}$ GeV. 
Thus, there is a reasonably wide range of values of $M_0$ (roughly between 
$10^{14}$ and $10^{15}$ GeV) where some particles are produced, but not 
enough for the backreaction to become an important issue.
 
If the condition~\leqn{bkimp} is satisfied, preheating proceeds in two stages. 
During the first stage, $q>q_2$ and the backreaction can be neglected. 
During the second stage, the backreaction has to be included. The 
``effective'' $q$-parameter at this stage is given by $q_{\rm eff} = 
\phi_0^2 /4\left< X^2 \right>$. This parameter is decreasing as more $X$ 
particles 
are created: approximate energy conservation implies $q_{\rm eff} \propto 
(\left< X^2 \right>)^{-2}$. Once $q_{\rm eff} \sim q_{min}$, particle   
production stops.

\subsection{Fermion Production.}
\label{nonSUSY2}

Several authors~\cite{Giudice, Greene, Peloso} have studied the fermion 
production during preheating using the Lagrangian
\beq
{\cal L}_f \,=\, (M_0 + g \phi) \bar{\psi} \psi.
\eeq{fermion}
They found that fermions as heavy as $10^{17}-10^{18}$ GeV could be 
produced. This is not surprising: with the Lagrangian~\leqn{fermion}, 
as long as the inflaton oscillation amplitude $\phi_0$ is greater than 
$M_0/g$, there is an instant of time during each oscillation when the 
effective mass 
of the fermion vanishes, and so the condition~\leqn{condition} is 
automatically violated at least for very large wavelengths. Thus, the
upper bound on $M_0$ is, for $g\sim 1$, of order of the initial value
of the inflaton, $0.3 M_p$. 

\section{Inflaton Interactions in Supersymmetric Models} 
\label{IInt}

Now, let us discuss how the above discussion of preheating generalizes to the 
supersymmetric case. Interactions of the inflaton with other fields are 
of two types: those coming from the Kahler potential and the superpotential. 
While the Kahler potential couplings such as $\int d^4\theta SX^\dagger X/
M_p$ can play a role in perturbative
reheating, they do not lead to resonant, 
non-perturbative particle production that we are interested in. Therefore, 
we will concentrate on superpotential interactions. The simplest 
supersymmetric generalization of~\leqn{scalar} is a model with two 
gauge-singlet chiral superfields, the ``inflaton'' $S$ and the ``matter 
field'' $X$, whose interactions are described by the 
superpotential\footnote{Supersymmetric models of chaotic inflation were 
constructed in Ref.~\cite{Murayama:xu}.}
\beq
W(S, X) = \frac{1}{2} m S^2 + \frac{1}{3}\eps S^3 + \frac{1}{2}(M + g S) X^2.
\eeq{super}
This is the most general renormalizable superpotential for two singlet 
fields with a discrete symmetry $X\rightarrow -X$. Without loss of generality,
we will choose the 
parameters $M$, $m$ and $g$ to be real and positive. The scalar potential  
is given by
\beq
V(S, X) = |mS + \eps S^2 + \frac{1}{2} gX^2|^2 + |M+gS|^2 |X|^2,
\eeq{spot} 
where $S$ and $X$ are complex scalar fields. Note that this potential
generically has four degenerate minima: $\{S=X=0\}$, $\{S=-m/\epsilon, X=0\}$
and $\{S=-M/g, X=\pm \sqrt{2(mM/g^2 - \epsilon M^2/g^3)} \}$. For $\eps=0$, 
the number of minima is reduced to three: $S=X=0$ and $\{S=-M/g, X =\pm 
\sqrt{2mM}/g.\}$ 
The presence of multiple, disconnected degenerate vacua, which will have 
important consequences for the dynamics of preheating, is completely generic 
in supersymmetric theories with multiple singlets. Indeed, the scalar 
potential in such theories has a form 
\beq
V(\phi_i) = \sum_{i=1}^n \left|\frac{\partial W(\phi_i)}{\partial \phi_i}
\right|^2.
\eeq{generalpot}
The vacua are found by solving a system of algebraic equations 
$\partial_i W(\Phi_i)=0, i=1 \ldots n$. For a renormalizable superpotential,
each of these equations is at most quadratic. Generically, the 
system has $2n$ different complex solutions, each of which corresponds to 
a vacuum with zero energy. One should keep in mind, however, that the 
degeneracy of these vacua is lifted by supersymmetry breaking unless the 
theory possesses additional discrete symmetries relating different vacua. 
(The model defined by~\leqn{super} happens to have such a symmetry, the 
$X \goesto -X$ reflection.) 

The expression \leqn{spot} can be thought of as the leading terms in 
the Taylor expansion of the full inflaton potential around $S=0$. This
potential is sufficient to describe the (p)reheating phase considered here,  
and is independent of the details of the slow roll phase of 
inflation.\footnote{We expect the qualitative features of our analysis to
hold in both chaotic inflation and new inflation models. Parametric 
resonance in supersymmetric hybrid inflation models was considered 
in~\cite{King}; however, it was subsequently shown that in these models
the inflaton typically decays via tachyonic 
preheating~\cite{tachyonic}, without developing a resonance.}
The only assumptions we make concern the values of the $S$ and 
$X$ fields at the end of the slow roll phase, which set the initial 
conditions for the process we are considering. Specifically, we assume 
that the initial value of the inflaton field $S_0$ is somewhat lower, but not 
too far from, the Planck scale $M_p$, while the initial value of $X$ is
zero. The latter assumption is reasonable since the mass of $X$ is large
(Planckian) during inflation. We treat the initial phase of the complex
field $S$ as arbitrary, which is a good approximation as long as the
potential in the phase direction is sufficiently flat during slow-roll
inflation. We will study the dependence of the results on this 
phase. 

The remaining freedom concerns the choice of parameters in the 
superpotential \leqn{super}. Throughout the analysis, we will assume that
$m \ll |S_0|$; for numerical estimates, we will use $m\sim 10^{13}$ GeV
and $|S_0| \sim M_p/3$ to facilitate the comparison with the 
non-supersymmetric analyses reviewed in Sec.~\ref{nonSUSY}. (These 
particular choices are motivated by the models of chaotic inflation
\cite{book}.) In this paper, we will primarily be interested in the
production of particles that are too heavy to be produced through 
thermal reheating or perturbative inflaton decay; therefore, we will
assume $M > m$. The size of parameter $\eps$ in chaotic 
inflation models is constrained to be at most about $10^{-6}$~\cite{Salopek}. 
Even such small coupling, however, can have 
profound effects on the inflaton evolution during the (p)reheating epoch. 
We will therefore consider two qualitatively different, physically 
interesting cases: $\eps=0$ and $\eps\sim 10^{-6}$. In the first case,
the complex phase of the inflaton is conserved during the 
evolution,\footnote{Strictly speaking, this is only true as long as the 
backreaction 
of the produced $X$ particles can be neglected - see Sec. \ref{SUSYsc}.}
and the motion of the inflaton is identical to the non-supersymmetric case. 
Still, there are some important differences as far as production of 
particles (especially fermions) is concerned. In the second case, the 
inflaton generally acquires angular momentum and moves along an elliptical 
spiral trajectory in the complex plane. As a result, the preheating process  
is quite different from the non-supersymmetric case.

\section{Scalar Particle Production in Supersymmetric Models.}
\label{SUSYsc}

In this section, we will consider the resonant production of scalar $X$ 
particles during the reheating era in the model defined by the 
superpotential~\leqn{super}. 

\subsection{$\eps=0$.} 
\label{sce0}

We start by examining the situation where the coefficient of the $S^3$ term
in the superpotential~\leqn{super} vanishes. As long as the backreaction of 
the $X$ particles can be neglected, the motion of the inflaton is 
described by the potential $V(S_1, S_2) = m^2|S|^2$. Since at 
the end of inflation the inflaton is slowly rolling, i.e. $\dot{S} \approx
0$, its subsequent evolution is given by
\beq
S(t) = S_0 e^{i\theta}\,\cos mt,
\eeq{osc}   
where $\theta$ is a constant phase, determined by the initial conditions
which we will assume to be completely random.
In an expanding universe, the amplitude of the oscillations $S_0$ becomes
time-dependent; assuming $S_0\approx M_p/3$ at $t=0$, after the first few 
oscillations $S_0 \approx M_p/3mt
\approx M_p/20N$, where $N$ is the number of oscillations that have been 
completed. The motion of the inflaton described by~\leqn{osc} is identical 
to the non-supersymmetric case, see Sec.~\ref{nonSUSY}.

The mass terms for the $X$ field can be read from \leqn{spot}:
\beq
\left( (M + gS_1)^2 + g^2 S_2^2 + gmS_1 \right) X_1^2 +   
\left( (M + gS_1)^2 + g^2 S_2^2 - gmS_1 \right) X_2^2 + 
2mgS_2 X_1 X_2,
\eeq{massesnd} 
where we have decomposed $X= X_1 + i X_2$ and $S= S_1 + iS_2$. 
Diagonalizing this mass matrix gives
\beq
\left( (M + gS_1)^2 + g^2 S_2^2 + gm |S| \right) \tilde{X}_1^2\,+\, 
\left( (M + gS_1)^2 + g^2 S_2^2 - gm |S| \right) \tilde{X}_2^2,
\eeq{masses} 
where $\tilde{X}_{1,2}$ are the mass eigenstates. (In what follows, we will
drop the tilde to avoid cluttering.)
The mode equations for the fields $X_{1,2}$ have the same form as \leqn{mode}, 
with the effective frequencies
\beq
\omega_k^2 = {{\bf k}^2\over a(t)^2} + (M + gS_1)^2 + g^2 S_2^2 \pm 
gm |S|,   
\eeq{omegac}
where the upper (lower) sign applies to the mode $X_1$ ($X_2$). This 
equation has some important differences from its non-supersymmetric 
counterpart, Eq.~\leqn{omega}. In particular, the right-hand side of 
\leqn{omegac} is {\it not} positive-definite. Since we are principally 
interested in heavy particle production, let us assume that $m \ll M$. 
Then, the right-hand side of~\leqn{omegac} for the field $X_2$ is always 
positive, whereas for the field $X_1$ it can become negative for 
sufficiently small ${\bf k}$. The region in the $S_1-S_2$ plane in which 
this is possible (the ``instability region'') is given by
\beq
(S_1-S_*)^2 + S_2^2 \leq {mM \over g^2},
\eeq{region}
where $S_* = -M/g$, and terms of order $m/M$ have been dropped. The 
instability region and a typical inflaton trajectory in the $\eps=0$ 
case are shown in Fig.~1. 

%--------------------------------------------------------------
\begin{figure}[t]
\begin{center}
\includegraphics[scale=0.6]{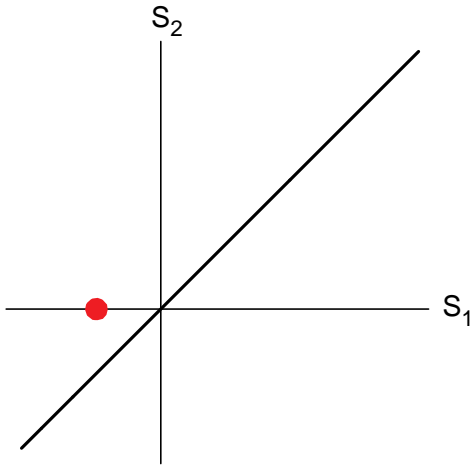}%
\caption{Motion of the inflaton in the $S_1-S_2$ plane in the case
$\eps=0$. The red circle indicates the instability region~\leqn{region}.}
\end{center}
\label{line}
\end{figure}
%-------------------------------------------------------------

Just like in the non-supersymmetric case, resonant production of $X$
bosons occurs whenever the condition~\leqn{condition} is violated. For 
the mode $X_2$, this can happen if
\beq
M^2 (\sin^2\theta + \frac{m}{M} \cos \theta) \lapproxeq \frac{1}{2}
gmS_0.
\eeq{x2prod}
For the mode $X_1$, the analysis depends on the phase $\theta$. If 
$\theta \gapproxeq \sqrt{m/M}$, the condition is similar to~\leqn{x2prod}:
\beq
M^2 (\sin^2\theta - \frac{m}{M} \cos \theta) \lapproxeq \frac{1}{2}
gmS_0.
\eeq{x1prod}
If, on the other hand, $\theta < \sqrt{m/M}$, the inflaton will pass 
through the instability region~\leqn{region}, provided that the 
oscillation amplitude is large enough, $S_0 > |S_*|$. During this passage,
the adiabaticity condition is bound to be violated, and production of
$X_1$ particles will occur. Thus, scalar particles with bare masses as high 
as $gS_0$ (i.e. up to about $10^{18}$ GeV, assuming $g \sim 1$) can in 
principle be produced in our model. However, production of particles 
substantially heavier than the limit indicated by the non-supersymmetric 
analysis of Sec.~\ref{nonSUSY}, Eq.~\leqn{massbound}, is only possible if 
$\theta < \sqrt{m/M}$. Since $m\ll M$, the range of $\theta$ for which 
this is true is quite small; in other words, production of superheavy 
scalar particles requires fine-tuned initial conditions.

The resonant production of $X$ particles can be described
analytically using the technique developed in Ref.~\cite{Linde1}. The
relevant calculations are presented in Appendix~\ref{app1}. For generic
initial conditions on the inflaton, the production rates are identical (up to 
${\cal O}(m/M)$ corrections) to those found in the non-supersymmetric case 
with the potential~\leqn{scalar} and $M_0 = M \sin\theta$. For special initial 
conditions, when the inflaton passes through the instability region, the 
production of $X_1$ particles is enhanced due to their locally imaginary
mass. As long as the inflaton oscillation amplitude is large, $S_0\gg |S_*|$,
the $X_1$ production rate is close to the rate estimated in \cite{Linde1} 
for non-supersymmetric particles with zero bare mass (see Eq.~\leqn{scalar} 
with $M_0=0$.) On the other hand, when $S_0 \sim S_*$, the $X_1$ production
rate can be dramatically enhanced.\footnote{Let us comment on the relation of 
our analysis in the case of ``special 
initial conditions'' to the tachyonic preheating scenario considered 
in~\cite{tachyonic}. In both cases, the field coupled to the inflaton 
develops an instability. However, in our case, the instability occurs only 
for a very short time period during each inflaton oscillation. As a result, 
the fractional energy loss by the homogeneous component of the inflaton in 
each passage is small. (This is true as long as $S_0\gg S_*$ -- a condition 
that is typically satisfied for a large number of inflaton oscillations at 
the beginning of the reheating era.) The backreaction of the produced 
$X$ particles does not become important until at least several inflaton 
oscillations have taken place, and the language of stochastic resonance 
is applicable.} These results are derived in Appendix~\ref{app1}.

To summarize our discussion so far, for generic initial conditions on
the inflaton field, $\theta \gg \sqrt{m/M}$, both $X_1$ and $X_2$ particles
are resonantly produced provided that
\beq
M \lapproxeq  \frac{1}{|\sin\theta|}\,\sqrt{\frac{gmS_0}{2}}.
\eeq{condSUSY} 
This condition is similar to the one obtained in~\cite{Linde1} for the 
non-supersymmetric case, Eq.~\leqn{massbound}. Indeed, if \leqn{massbound}
is satisfied, preheating occurs regardless of the initial phase of the
inflaton field. Preheating of particles with higher bare masses is possible,
but requires special initial conditions. The absolute upper bound on $M$ is
achieved for the case when the inflaton is real ($\theta=0$) and is of
order $gS_0 \sim 10^{18}$ GeV for $g\sim1$. Moreover, for the same special 
initial conditions, the production of $X_1$ bosons can be significantly 
enhanced.

The above discussion neglected the backreaction of the produced $X$ 
particles on the motion of the inflaton. Let us now include this effect.
First, consider generic initial conditions, $\theta \gg \sqrt{m/M}$. In
this case, the rough estimate of the produced $X$ number density 
is the same as in the non-supersymmetric case discussed in 
Sec.~\ref{nonSUSY}, with the replacement $M_0 \rightarrow M \sin\theta$. 
Thus, the backreaction can be neglected for $M\sin \theta \gapproxeq
8m \approx 8 \times 10^{13}$ GeV. For smaller values of $M\sin \theta$,
the $X$ particles are still being produced at the time when their 
backreaction becomes important. Including it leads to an effective
inflaton potential of the form
\beq
V(S) = g^2 \xsq\left( (S_1-S_*)^2 + S_2^2 \right) + m^2|S|^2.
\eeq{effectif}
The phase of $S$ is not conserved by the first term in this potential, and 
will start changing (tending towards zero) when the backreaction becomes
important. As a result, $S_1$ and $S_2$ oscillations will be out of
phase. The resonant $X$ particle production will quickly terminate,
since it is possible only when $S_1$ and $S_2$ are {\it both} close to 
the minimum of the potential. For the modes $X_{1,2}(k)$ with large occupation
numbers, the subsequent evolution can be considered classically. However, 
it is still very complicated, with all the four coupled fields $X_{1,2}$ 
and $S_{1,2}$ undergoing large oscillations in the potential
\beq
V(S, X) \approx g^2 |X|^2\left( (S_1-S_*)^2 + S_2^2 \right) + \frac{1}{4}
g^2 |X|^4, 
\eeq{effectif1}
where we have assumed $g^2 \xsq \gg m^2$.
The oscillations are damped by Hubble friction. When the oscillation 
amplitudes become sufficiently small, $g^2 \xsq<m^2$ or $\left< (S_1-S_*)^2 
\right> < mM$, the terms in the scalar potential which were neglected
in Eq.~\leqn{effectif1} become important. It is at this stage that the
system chooses which of the three vacua will be the final point of the
evolution. This is a very important question. If the vacua with non-zero 
$X$ and $S$ are preferred, the phenomenology will crucially depend on the
exact nature of the $X$ field. For example, if the $X$ field carries $R$ 
parity, these vacua are clearly undesirable. Moreover, in our model there 
are two vacua with non-zero $X$. The $X$ fields whose evolution is      
considered here are position-dependent, since modes in a wide range of 
wavelengths obtain large occupation numbers during preheating. It is 
therefore very likely that different vacua will be chosen in different 
spatial regions, leading to formation of domain walls. (This non-thermal
defect production mechanism is similar to the one considered, for example, 
in~\cite{phase}.) Because of the $Z_2$ symmetry relating the two 
vacua, they 
will remain degenerate even after supersymmetry is broken, and the domain 
walls of our model are stable.\footnote{In more general models, 
supersymmetry breaking can lift the vacuum degeneracy, leading to unstable 
domain walls. In this case, the cosmological bounds could be avoided, but
a more careful examination of the domain wall dynamics is necessary.} 
Therefore, the model is only consistent with standard cosmology  
if the system ends up in the vacuum $S=X=0$ throughout the space. It would 
be very interesting to 
study the evolution in more detail and understand which vacuum is chosen for 
various values of the parameters and initial conditions. This will likely 
require a numerical study using the methods developed in~\cite{KT, FT}.      
             
For special initial conditions, $\theta \lapproxeq \sqrt{m/M}$, the 
backreaction is expected to be very important. Indeed, the $X_1$ particle 
production rate in the instability region~\leqn{region} is at least as high
as the rate in the non-supersymmetric, $M_0=0$ case considered 
in~\cite{Linde1}. Based on the analysis of~\cite{Linde1}, we conclude that 
the backreaction of $X_1$ particles will become important after just a few 
passages of the inflaton through the instability region. Moreover, in this 
case the $X_1$ production will continue even after the backreaction becomes 
important, and only stop when $\xsq \sim \left<(S_1-S_*)^2 \right>.$ After 
that, the evolution of the modes $X_1(k)$ with large occupation numbers
can be described as classical motion of the fields $X_1$ and $S_1$ in the 
potential~\leqn{effectif1}, with $X_2 \approx S_2 \approx 0$, damped by  
Hubble friction. A quick estimate suggests that the evolution is more 
likely to end in the vacua with non-zero $X$, leading to domain wall 
production.    

Summarizing, we have found that for generic initial conditions,
$\theta \gg \sqrt{m/M}$, and sufficiently high masses, $M\sin\theta 
\gapproxeq 8m$, preheating shuts down before the backreaction of the produced 
particles on the inflaton motion becomes important, and the picture is quite 
similar to the non-supersymmetric case studied in~\cite{Linde1}. On the
other hand, for $M\sin\theta \lapproxeq 8m$ the backreaction becomes an
important effect. Once it is taken into account, the evolution of the system 
is rather more involved than in the non-supersymmetric case, and may lead
to unacceptable consequences such as production of domain walls. Clearly,
further investigation of this situation is necessary. Furthermore, for 
special initial conditions, $\theta \lapproxeq \sqrt{m/M}$, we found that 
the upper bound on the masses of the $X$ bosons that can be reheated is
substantially higher than in the non-supersymmetric case. In this case, 
backreaction is expected to be important, and it seems likely that domain 
walls are produced at the final stages of the process. Again, a more detailed 
numerical study is necessary to check this conclusion. 

\subsection{$\eps\not=0$.}
\label{SCeps}

The cubic term $S^3$ in the superpotential \leqn{super} has important 
consequences for the motion of the inflaton after the end of the slow-roll 
inflation and, therefore, for preheating. In the presence of this term, the
motion of the inflaton is governed (neglecting backreaction) by the
potential  
\beq
V(S_1, S_2) = m^2(S_1^2+S_2^2) + 2\eps m(S_1^2+S_2^2)S_1 +
\eps^2 (S_1^2+S_2^2)^2.  
\eeq{spoteps}
The second term in this potential violates the symmetry $S \goesto e^{i\theta}
S$. Due to this term, the motion of the inflaton for generic initial 
condition is no longer decribed by \leqn{osc}, but is much more 
complicated. We will restrict our analysis to the case when the terms 
proportional to $\eps$ in \leqn{spoteps} are of the same order as
the first term at the end of the slow-roll inflation. (With our standard 
assumption $m\sim10^{13}$ GeV this implies $\eps\sim10^{-6}$, close to 
its upper bound in chaotic inflation models~\cite{Salopek}.) The evolution of 
the inflaton during the reheating era can be broken into two stages. During
the first stage, the effect of the cubic and quartic terms in the
potential \leqn{spoteps} is significant, and the motion is complicated.
The potential \leqn{spoteps} has two vacua: $S=0$ and $S=-m/\eps$. 
Because of the Hubble friction, the energy of the system is decreasing and
it will eventually settle down into one of these vacua. Let us assume that 
the system ends up in the vacuum with $S=0$. (This is always the case when 
$m/\eps \gg |S_0|$. For the case considered here, $m/\eps\sim|S_0|$, the 
choice of vacuum depends on 
the initial conditions; domains of attraction of the two vacua have more or 
less the same size.) In this case, the average magnitude of the inflaton field 
$|S|\equiv\sqrt{S_1^2+S_2^2}$ decreases with time. Eventually the
quadratic term in \leqn{spoteps} becomes dominant, and the second stage of
the inflaton evolution begins. At this stage, the 
potential can be approximated by setting $\eps=0$. However, the problem 
does not reduce to the case studied in the previous subsection, since the 
inflaton field generically has non-zero angular momentum. Its motion can be 
described as independent, {\it out-of-phase} oscillations of $S_1$ and 
$S_2$ fields:
\beqa
S_1 &=& S_{1,0}(t) \sin mt, \CR    
S_2 &=& S_{2,0}(t) \sin (mt + \delta),
\eeqa{ellips}
where $S_{1,0}$ and $S_{2,0}$ are slowly decreasing with time as a result of
the Hubble expansion. Thus, during the second stage, the inflaton is 
approaching the minimum of its potential along an elliptical spiral. As we
show in Appendix~\ref{appb}, with the assumptions made here, the ellipticity 
of the orbit is generally of order one. A typical trajectory of the inflaton
is sketched in Fig.~2.

%--------------------------------------------------------------
\begin{figure}
\begin{center}
\includegraphics[scale=0.8]{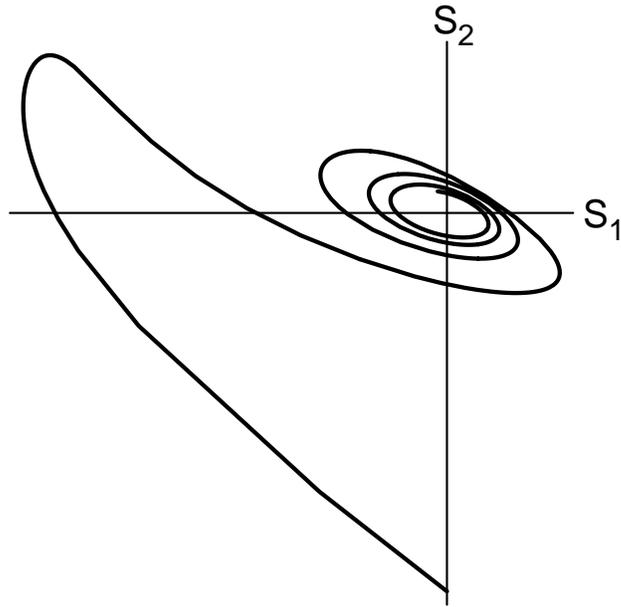}%
\caption{A typical trajectory of the inflaton in the case $\eps\not=0$. 
The first part of the trajectory depends on the initial conditions and
does not have a universal shape. The second part of the trajectory,
corresponding to $\eps |S| \ll m$, is an elliptical spiral (unless the 
phase of $S$ vanishes at the end of inflation.)}

%The
%circle indicates the instability region~\leqn{region}. The figure is not to 
%scale: for typical parameters of our model, the instability region is much 
%smaller and much closer to the origin than shown here.}
\end{center}
\label{spi}
\end{figure}
%-------------------------------------------------------------

The resonant production of $X$ particles is only possible if the 
non-adiabaticity condition \leqn{condition} is satisfied. For low-momentum 
modes, the frequency can be approximated by
\beq
\omega^2 = (M + gS_1)^2 + g^2S_2^2 \mp gm|S|.
\eeq{freq}
During the first stage of the reheating era, the resonant $X$ production  
is very unlikely. Indeed, approximate energy conservation implies that the 
velocity of the inflaton satisfies $|\dot{S}| \lapproxeq m|S_0|$. Since 
$m \ll g|S_0|$, the non-adiabaticity condition can be satisfied only if 
the frequency $\omega$ is much smaller than its typical value, which is order 
$g|S_0|$. If $M \ll g|S_0|$, this can only occur when both $S_1$ and $S_2$ 
are close to zero, $|S|\lapproxeq M$; if $M \sim g|S_0|$, this requires 
that the inflaton pass through (or close to) the instability region 
\leqn{region}. The area of these ``production regions'' in the $S_1-S_2$ plane
is quite small compared to the range of the inflaton motion during the 
first stage. Since the inflaton motion is highly non-periodic, it is very 
unlikely that it will pass through the ``production regions'' repeatedly, as 
required for the buildup of the number density in the resonant production 
scenario.  
 
Now, let us consider the second stage of the preheating era, when the motion
of the inflaton is approximately described by \leqn{ellips}. For simplicity,
let us set $S_{1,0}=S_{2,0}$ and $\delta=\pi/2$, so that the orbit is a 
circular spiral. (The physical results of our analysis are insensitive to
these assumptions.) In this case, the condition \leqn{condition} can be
studied analytically. The bottomline is that the $X$ particle production only 
occurs when the inflaton passes through the ``production region'' in the
$S_1-S_2$ plane. This production region coincides (up to order-one factors)
with the instability region defined by Eq.~\leqn{region}. This is not 
surprising: it is clear that the adiabaticity condition is always violated 
in the instability region. In the $\eps=0$ case studied in the previous 
subsection, we found that $X$ particles can be produced even without the 
inflaton passing through the instability region as long as their mass is 
sufficiently low, see Eq.~\leqn{condSUSY}. This production occurs twice per 
oscillation period when the inflaton passes the origin, provided that its 
velocity $|\dot{S}|\sim mS_0$ is greater than (roughly) $M/g$. In the case 
of spiral motion considered here, however, the inflaton does not get close 
to the origin until it loses most of its energy. Thus, the only possible 
production mechanism in this case is by passing through the instability 
region. 

%--------------------------------------------------------------
\begin{figure}
\begin{center}
\includegraphics[scale=0.6]{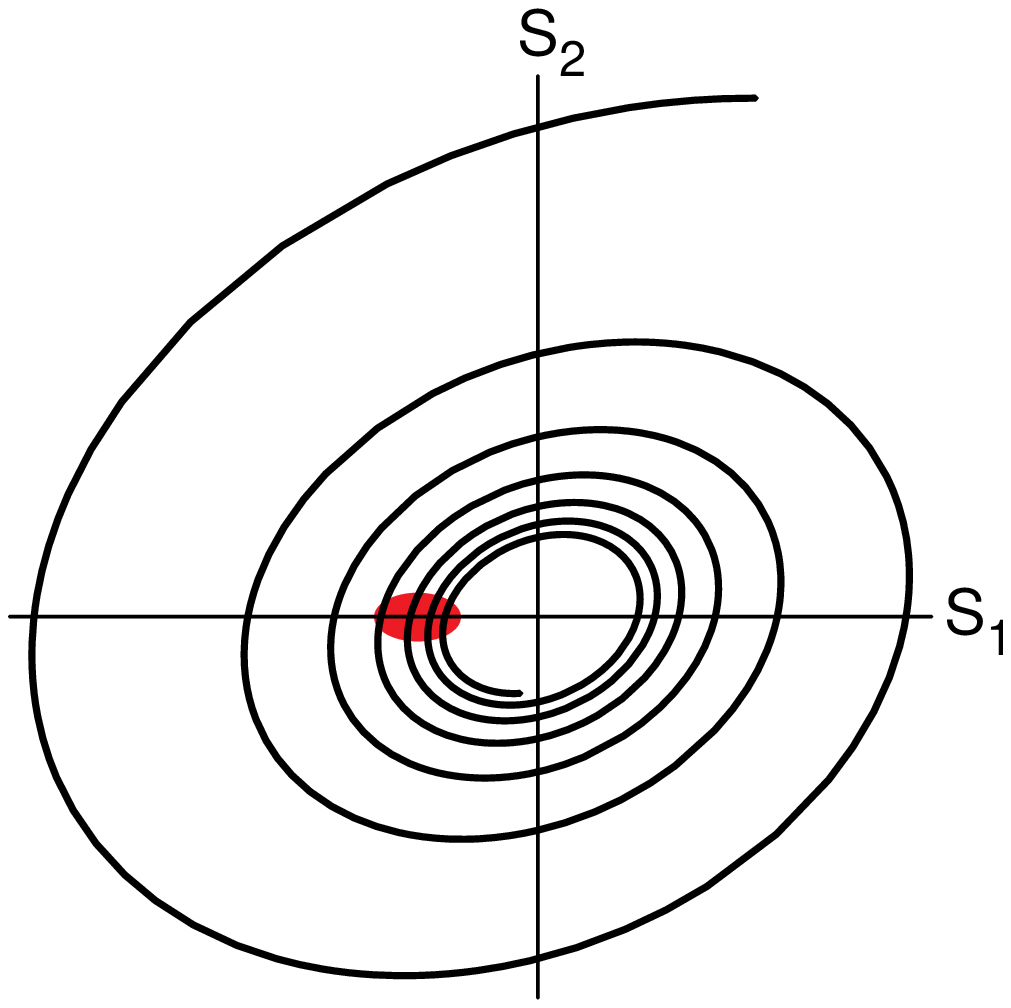}%
\hskip1cm
\includegraphics[scale=0.6]{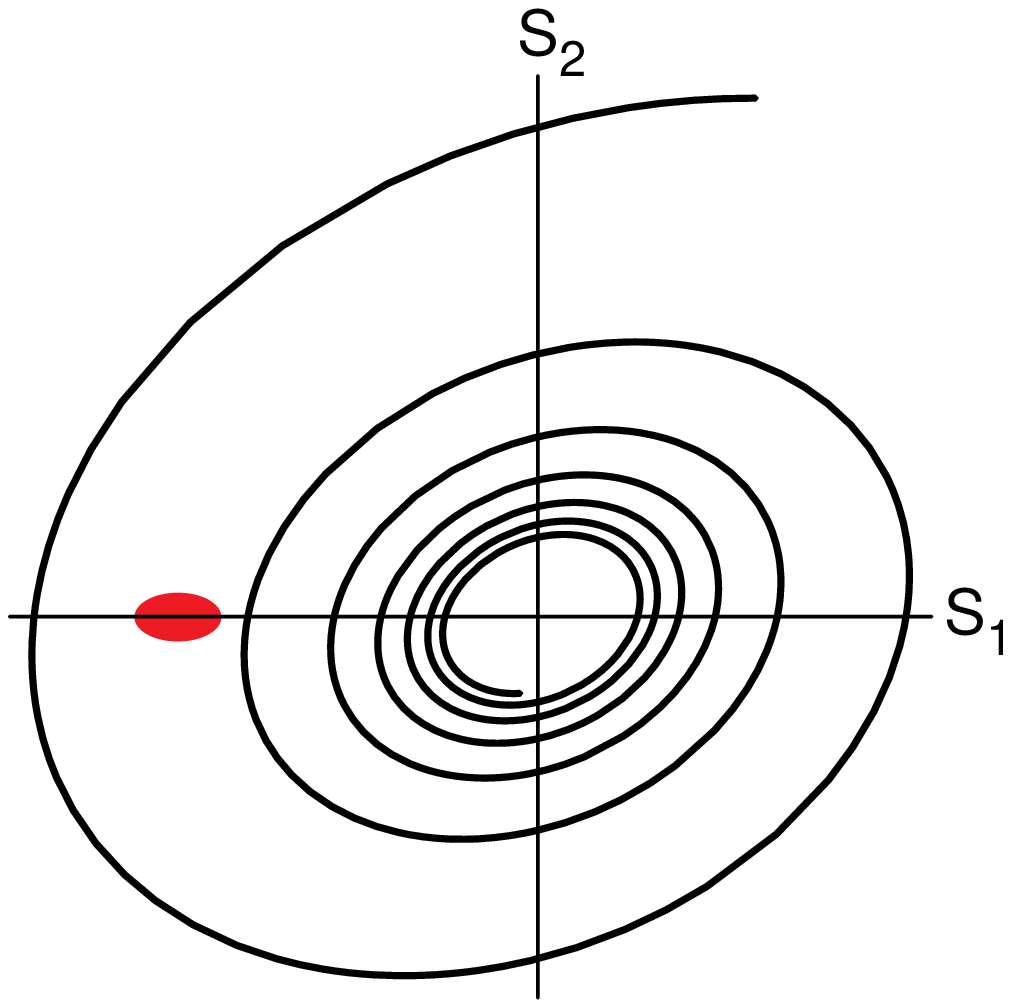}%
\caption{Depending on the step of the spiral, the inflaton either passes 
through the instability region (left panel) or misses it (right panel).}
\end{center}
\label{spii}
\end{figure}
%-------------------------------------------------------------
  
The inflaton will pass through the instability region regardless of the 
initial conditions only if the step of its spiral trajectory $\delta$ at the
time when $|S| \approx S_*$ is much smaller than the extent of the instability
region along the $S_1$ axis, $\Delta S \approx \sqrt{mM}/g$. (This is 
illustrated in Fig.~3.) Estimating 
$\delta \approx S_*^2/M_p$, we find that this requirement implies an
upper bound on the bare mass of the $X$ particles that can be produced:
\beq
M \lapproxeq (g^2 m M_p^2)^{1/3}.
\eeq{newbound}
This bound is somewhat higher than the corresponding bound in the 
non-supersymmetric case, Eq.~\leqn{massbound}, and the bound for generic 
initial conditions in the case $\eps=0$, Eq.~\leqn{condSUSY}. For our
standard parameter values, $g\sim 1$ and $m\sim10^{13}$ GeV, 
Eq.~\leqn{newbound} implies $M \lapproxeq10^{17}$ GeV. 

Particle production by a spiralling inflaton passing through the 
instability region is considered quantitatively in Appendix~\ref{app1}. The 
production is extremely efficient: in fact, the backreaction of the produced 
$X_1$ particles on the inflaton may become important already after two or 
three production events, terminating the process. This explosive production 
is analogous to the so-called tachyonic preheating~\cite{tachyonic}, which 
frequently occurs in hybrid inflation models. The subsequent evolution of
the system can be studied classically, but is very complicated due to a large
number of fields involved. It seems likely that this evolution will involve 
domain wall formation. 

In summary, we have found that if the cubic term is present in the inflaton
superpotential, the dynamics of preheating is qualitatively different from
the non-supersymmetric case. If the condition~\leqn{newbound} is satisfied,
non-perturbative particle production occurs when the inflaton trajectory is
spiral, and does not require fine-tuned initial conditions. Production in 
this case tends to be explosive, with the inflaton losing a large part of its 
energy already during the first two or three production events. For larger 
values of $M$, particle production may in principle occur during the initial, 
irregular phase of the inflaton motion; however, for generic initial 
conditions, this possibility is highly unlikely.  

\section{Fermion Production in Supersymmetric Models.}
\label{SUSYf}

Preheating of fermionic particles in supersymmetric models has 
important differences from the non-supersymmetric case reviewed in
Sec.~\ref{nonSUSY2}. The matter chiral superfield $X$ contains two
two-component (Weyl) spinors, which we will denote by $\psi_L$ and
$\psi_R$. The interactions of these fields follow from the 
superpotential~\leqn{super}:  
\beq
{\cal L}_f = -(M+gS)\psi_R^\dagger \psi_L -\,(M+gS^*)\psi_L^\dagger \psi_R.
\eeq{fermionsusy}
Combining the fields $\psi_L$ and $\psi_R$ into a four-component (Dirac) 
spinor $\Psi = (\psi_L, \psi_R)^T$, Eq.~\leqn{fermionsusy} can be 
rewritten as
\beq
{\cal L}_f = -(M+gS_1)\,\bar{\Psi}\Psi +\, igS_2 \bar{\Psi} \gamma^5 \Psi,
\eeq{fermionsusy1}
where $S=S_1+iS_2$, $\bar{\Psi}=\Psi^\dagger \gamma^0$, and we have chosen 
the Weyl basis in which $\gamma^5=$ diag$(-1,-1,1,1).$ 
Comparing Eq.~\leqn{fermionsusy1} with its non-supersymmetric counterpart, 
Eq.~\leqn{fermion}, indicates an important difference: the appearance of an 
additional, pseudoscalar mass term in the supersymmetric case. This term 
plays a significant role in the analysis of preheating. 

The evolution of the fermion wavefunctions is governed by the Dirac equation
of the form $i \partial_t \Psi(t) = ${\bf A}$(t) \Psi(t)$, where
\beq
{\rm {\bf A}}(t) = \left( \begin{array}{cc} 
  \pm k & M + gS(t) \\ 
  M + gS^*(t) & \mp k 
       \end{array} 
\right).
\eeq{dirac} 
In this equation, $k$ is the momentum of the mode and we are working in the
helicity basis, $i\sigma^i\partial_i\psi=\pm k\,\psi$. The frequencies of
the fermionic modes are given by the eigenvalues of the matrix {\bf A}, 
which are in general time-dependent: $\omega = \pm \sqrt{k^2+|M+gS(t)|^2}$. A 
necessary condition for the adiabatic evolution of the system is 
$\dot{\omega} \ll \omega^2$; when this condition is violated, 
the evolution is non-adiabatic and particle production is expected to occur.
Below, we will use this condition to obtain a qualitative picture of 
fermion preheating in supersymmetric models.

\subsection{$\eps=0$}

As expected in a supersymmetric theory, the analysis of fermionic preheating 
in this case is very similar to the scalar preheating analysis of
Sec.~\ref{sce0}. For $m \ll M \lapproxeq \sqrt{mgS_0/2}$, fermions can be 
produced for any value of the inflaton phase, whereas for $\sqrt{mgS_0/2}
\lapproxeq M \lapproxeq gS_0$, the production occurs if this phase 
satisfies 
\beq
|\sin\theta| \lapproxeq  \frac{1}{M}\,\sqrt{\frac{gmS_0}{2}}.
\eeq{phase} 
The production of very heavy fermions (up to $M\sim gS_0$)
is possible,\footnote{Naively, the expression \leqn{phase} seems to imply 
that infinitely heavy fermions can be produced for $\theta=0$. This is of 
course incorrect; the bound \leqn{phase} is not applicable for $\theta 
\lapproxeq \sqrt{m/M}.$ The correct bound in that case is $M \lapproxeq
gS_0$.} but only if the initial phase of the inflaton is fine-tuned 
to be close to 0. (This fine-tuned case is identical to the non-supersymmetric
model of Sec.~\ref{nonSUSY2}.) For generic initial conditions, the upper bound 
on the mass of the fermions that can be effectively preheated is of order 
$\sqrt{gmS_0/2}$, well below the non-supersymmetric estimate. 

The condition $\dot{\omega} \ll \omega^2$ is necessary for adiabaticity, 
but it is not sufficient. Particle production could occur even when 
$\dot{\omega}=0$
if the eigenvectors of the matrix {\bf A} have strong time-dependence. For
example, a rapid change of sign of the effective mass $M+gS$, even without
a change in its magnitude, would lead to particle production. In the case
$\eps=0$, however, it is possible to show that the eigenvector rotation is 
adiabatic whenever the condition $\dot{\omega} \ll \omega^2$ is satisfied.    
Thus, fermion production does not occur outside of the parameter regions 
discussed in the previous paragraph.

It is important to keep in mind that
in supersymmetric models, fermion and boson production cannot be considred
separately: the fermion and scalar masses and their couplings to the inflaton 
are all expressed in terms of just three superpotential parameters, $M$, $m$ 
and $g$. As we emphasized in Section~\ref{sce0}, some regions of this 
parameter space may not lead to consistent cosmology, for example due to 
non-thermal production of topological defects. If this is the case, additional 
restrictions would be imposed on the masses of the fermions that can be 
preheated. In particular, the production of superheavy fermions, $M >
\sqrt{gmS_0}$, requires a passage of the inflaton through (or close to) the
instability region, which results in enhanced scalar particle 
production and may be accompanied by domain wall formation.

Even if the fermion is light enough to be preheated, we expect the number of 
produced particles for generic initial conditions to be smaller than in the 
non-supersymmetric case. Indeed, the effective frequency of the modes in 
the supersymmetric case is limited from below by $M\sin\theta$, leading to
a suppression in the efficiency of particle production for non-zero $\theta$. 
The magnitude of this suppression could be estimated by a numerical 
integration of Eq.~\leqn{dirac}. Such an investigation is outside the 
scope of this paper.

\subsection{$\eps\not=0$.}

Again, the analysis is completely analogous to the corresponding
analysis of the scalar production presented in Section~\ref{SCeps}. 
There are two possibilities for fermion production. First, it can 
occur when the inflaton passes close to the origin in the $S_1-S_2$ 
plane, with a sufficiently large velocity. The second possibility is when 
the inflaton passes through the ``instability region'' around the point
$S_1=-M/g, S_2=0$. (The fermion effective mass vanishes at that point.)
To obtain resonantly enhanced production, the inflaton should pass 
through either one of these special regions repeatedly.  
 
For non-zero $\eps$, the reheating epoch can be subdivided into two 
stages. During the first stage, the inflaton motion is highly non-periodic 
and irregular (see Fig.~2.) Therefore, repeated passage through either one 
of the two relatively small production regions is unlikely. During the 
second stage, the inflaton approaches its minimum along an elliptically 
spiral trajectory. Fermion production is possible if this trajectory 
passes through the instability region, given by \leqn{region}. As shown
in Section~\ref{SCeps}, this puts an upper bound on the bare mass of the
particles that can be preheated, $M < (g^2 m M_p^2)^{1/3}.$ However, if
this condition is satisfied, the production of scalars in the instability 
region is explosive, generically leading to undesireable consequences 
such as domain wall formation. Thus, obtaining a cosmologically consistent 
scenario of fermionic preheating in a supersymmetric model with $\eps\not=0$ 
is difficult. Given this conclusion, we will not attempt to calculate the 
number density of fermions produced in this case.   

\section{Conclusions.}
\label{Conc}

In this paper, we have studied the implications of supersymmetry for the 
inflaton dynamics in the reheating era and the process of preheating. The 
emerging picture shows important differences from the previous, 
non-supersymmetric analyses. This is due to several robust, generic features 
of supersymmetric theories. First, all the scalars in supersymmetric theories 
are necessarily complex, meaning that an extra degree of freedom, the inflaton 
phase, has to be considered. Second, the couplings between the inflaton and 
the ``matter'' fields (the fields that are being preheated) are constrained by 
holomorphy. Third, supersymmetric models usually possess multiple degenerate 
vacua, which can become relevant (and lead to cosmological difficulties) if 
the density of particles produced during preheating is large. Finally, the 
preheating of bosons and fermions in a supersymmetric model cannot be 
considered in isolation: both kinds of particles are necessarily present, and 
their couplings are related.    
 
Several applications of preheating were suggested in the literature, and have 
been argued (using non-supersymmetric models) to be successful. Do these 
optimistic conclusions survive in supersymmetric theories? Our analysis 
demonstrates that the answer to this question is not straightforward. For 
example, consider a scenario of leptogenesis where heavy ($10^{14}$ GeV) 
right-handed neutrinos are produced via preheating. This process is described 
by our toy model, Eq.~\leqn{super}, if we identify $X$ with the right-handed 
neutrino chiral superfield. Generically, the parameter $\eps$ is non-zero, and 
the analysis of Sections 4.2 and 5.2 applies. Eq.~\leqn{newbound} shows that 
sufficiently heavy right-handed neutrinos can indeed be produced. However, at 
the same time, their scalar superpartners are produced explosively (see 
Section 4.2 and Appendix A.2.) As a result, the system is likely to end up in 
the vacua with non-zero right-handed sneutrino vacuum expectation value, 
either in some regions of space or everywhere. Both possibilities are 
phenomenologically disastrous. Therefore, the overall success of the 
leptogenesis scenario is not guaranteed, and may involve significantly more 
parameter fine-tuning and/or additional model building than previously thought.

Our analysis is only the first step in the quantitative investigation of 
preheating after inflation in supersymmetric models. It needs to be extended 
in several directions. For example, in many cases, preheating is so efficient 
that the backreaction of matter should be taken into account at the end of 
this process. We have neglected this effect throughout the 
analysis, giving only qualitative estimates for the parameter ranges over 
which it becomes important. Studying the evolution of the inflaton-matter 
system after the end of preheating, including the backreaction, is crucial 
since it will allow one to answer the all-important questions about the vacuum
selection and domain wall formation. Also, we have not attempted to describe 
quantitatively the regions of parameter space that allow for successful 
applications of preheating, e.g. to leptogenesis or WIMPZILLA production. 
Further work in this direction is necessary.

%%%%%%%%%%%%%%%%%%%%%%%%%%%%%%
\vskip0.7cm
\noindent
{\bf Acknowledgments}\\
\noindent

We thank Gary Felder and Patrick Greene for fruitful discussions. The authors 
are supported by the Director, Office of Science, Office of High
Energy and Nuclear Physics, of the U. S. Department of Energy
under Contract DE-AC03-76SF00098, and by the National Science
Foundation under grant PHY-00-98840.

\appendix
\section{Analytic Description of Scalar Preheating.}
\label{app1}

In this appendix, we will present a semi-analytic treatment of the 
production of scalar particles $X$ during inflaton oscillations. The 
qualitative aspects of this phenomenon have been discussed in 
Section~\ref{SUSYsc}. The calculation presented here is very similar to 
the one performed in~\cite{Linde1} for the case of real scalars. However,
our results are more general; in particular we will consider the case when 
the inflaton field passes through the instability region \leqn{region}, 
and show that this generally leads to enhanced particle production.

\subsection{$\eps=0$.}

Let us first concentrate on the case $\eps=0$, discussed in 
Section~\ref{sce0}. Consider production of the quanta of the field $X_1$. 
Defining $\Theta(t) \equiv a^{-3/2}(t)X_1(t)$, the mode equation reads
\beq
\ddot{\Theta}_k + \omega_k^2 \Theta_k = 0
\eeq{ap:mode}
with the effective frequency
\beq
\omega_k^2 = \frac{{\bf k}^2}{a^2(t)} + g^2(S_1-S_*)^2 + g^2 S_2^2 -gm|S|
+ \Delta,
\eeq{ap:omega} 
where $S_* = -M/g$, and terms of order $m/M$ are neglected. The correction 
term $\Delta=-\frac{3}{4}(\dot{a}/a)^2-\frac{3}{2}(\ddot{a}/a)$ is also
typically very small and will be neglected throughout the analysis. 

The motion of the inflaton field for $\eps=0$ is described by Eq.~\leqn{osc}. 
Particle production occurs only when the non-adiabaticity condition 
\leqn{condition} is satisfied. For generic initial conditions, this
requires $M \ll S_0$, see Eq.~\leqn{condSUSY}, and production occurs
twice during each inflaton oscillation cycle, when $S$ is close to the origin.
For special initial conditions production occurs when the inflaton passes
through the instability region \leqn{region}. In both cases,  
each of the production periods is short compared to the inflaton 
oscillation period. This leads to the following approximation. Let $t_j$ 
denote the moments of time when the $S$-dependent terms in $\omega_k$ are
minimized:
\beq
\cos mt_j = -{M\cos\theta \over gS_0}.
\eeq{ap:times}   
Particle production occurs at $t_j-\Delta t \lapproxeq t \lapproxeq t_j+
\Delta t$, where $m\Delta t \ll 1$. Outside of these time intervals, the 
mode equation \leqn{ap:mode} can be solved by making the adiabatic 
approximation. For $t_{j-1}+\Delta t \lapproxeq t \lapproxeq t_j-\Delta t$,
we write
\beq
\Theta_k(t) = \frac{\alpha^j_k}{\sqrt{2\omega}}\,\exp\left(-i\int_0^t \omega 
dt \right) \, + \, \frac{\beta^j_k}{\sqrt{2\omega}}\,\exp\left(+i\int_0^t 
\omega dt \right),
\eeq{ap:adiabat}
where the coefficients $\alpha^j_k$ and $\beta^j_k$ are constant. In the 
next adiabatic time period, $t_j+\Delta t \lapproxeq t \lapproxeq t_{j+1}-
\Delta t$, the expression for $\Theta_k(t)$ has the form
\beq
\Theta_k(t) = \frac{\alpha^{j+1}_k}{\sqrt{2\omega}}\,\exp\left(-i\int_0^t 
\omega dt \right) \, + \, \frac{\beta^{j+1}_k}{\sqrt{2\omega}}\,\exp\left(+i
\int_0^t \omega dt \right).
\eeq{ap:adiabat1}
With our normalization, the coefficients $\alpha$ and $\beta$ are just  
the coefficients of the Bogolyubov transformation of the creation and 
annihilation operators. The comoving density of $X_1$ particles at time $t$
is given by 
\beq  
n_{X_1}(t) = {1 \over 2\pi^2a^3} \int_0^\infty dk k^2 |\beta_k(t)|^2.
\eeq{ap:density}
When the adiabaticity condition holds, the number density in \leqn{ap:density}
is constant, since $\beta_k$ does not change. When the condition is violated,
the number density can change: in general $\beta_k^{j+1} \not=
\beta_k^j$. Thus, to estimate the change in particle density during a 
single production event, we need to find the Bogolyubov transformation 
matrix {\bf B} corresponding to this event:
\beq
\left(\begin{array}{c} \alpha_k^{j+1} \\ \beta_k^{j+1} \end{array}\right)
\,=\,{\rm {\bf B}}_k^j \,
\left(\begin{array}{c} \alpha_k^j \\ \beta_k^j \end{array}\right).
\eeq{ap:Bogol}
To achieve this, we perform a Taylor expansion of the effective frequency 
\leqn{ap:omega} for $t \approx t_j$:
\beq
\omega_k^2 = \frac{{\bf k}^2}{a^2(t_j)} + \omega_{k,min}^2 + g^2 |\dot{S}
(t_j)|^2 (t-t_j)^2,
\eeq{ap:taylor}
where $\omega_{k,min}^2 = M^2 \sin^2 \theta - mM \cos\theta$. For 
generic initial conditions, the production occurs when $S\approx0$, and 
therefore $|\dot{S}(t_j)|\approx mS_0$; whereas for special initial 
conditions, we have $|\dot{S}(t_j)|\approx m\sqrt{S_0^2-S_*^2}$. Defining 
$k_* = \sqrt{g|\dot{S}(t_j)|}$, $\tau = k_*(t-t_j)$, $\kappa = 
{\bf k}/a(t_j)k_*$ and using \leqn{ap:taylor}, the mode equation at 
$t \approx t_j$ becomes
\beq
{d^2\Theta_k \over d\tau^2} + (\lambda_k^j + \tau^2) \Theta_k = 0
\eeq{ap:modeT}
where
\beq
\lambda_k^j = \kappa^2 + \frac{1}{k_*^2}\left(M^2 \sin^2 \theta - mM
\cos\theta \right).
\eeq{ap:lambda}
Note that the parameter $\lambda_k^j$ can be negative for modes with low
momenta if the inflaton passes through the instability region 
\leqn{region}. The exact analytic solution of \leqn{ap:modeT} is given by
parabolic cylinder functions~\cite{GR}:
\beq
\Theta_k(\tau) = C_1 D_p((1+i)\tau) + C_2 D_p(-(1+i)\tau),  
\eeq{ap:soln}
where $p=-(1+i\lambda_k^j)/2$ and $C_{1,2}$ are arbitrary coefficients. The
form \leqn{ap:soln} is valid for both positive and negative $\lambda$.
The asymptotic forms of $\Theta_k(\tau)$ for $\tau\goesto-\infty$ and
$\tau\goesto+\infty$ should match the adiabatic solutions before and after 
the production event, Eqs.~\leqn{ap:adiabat} and~\leqn{ap:adiabat1}
respectively. Performing this matching allows us to express the four
Bogolyubov coefficients, $\alpha_k^{j,j+1}$ and $\beta_k^{j,j+1}$, in terms
of only two constants $C_1$ and $C_2$, and therefore to find the
transformation matrix {\bf B}$_k^j$ in \leqn{ap:Bogol}. The result is
\beq
{\rm {\bf B}}_k^j \,=\,
\left( \begin{array}{cc} 
  \sqrt{1+e^{-\pi\lambda}}e^{-i\varphi} & ie^{-\pi\lambda/2+2i\theta_k^j} \\ 
      -ie^{-\pi\lambda/2-2i\theta_k^j} & \sqrt{1+e^{-\pi\lambda}}e^{i\varphi} 
       \end{array} 
\right),
\eeq{ap:matrix}   
where $\lambda \equiv \lambda_k^j$, $\theta_k^j = \int_0^{t_j} 
\omega(t)dt$ is the phase accumulated by the moment $t_j$, and the angle 
$\varphi$ is given by
\beq
\varphi = \arg \Gamma(\frac{1+i\lambda}{2}) \,+\,\frac{\lambda}{2}\,\left(
1+\log \frac{2}{\lambda} \,\right).
\eeq{ap:phase} 
The results in section VII of the paper~\cite{Linde1} correspond to setting
$\lambda=\kappa^2$ in these formulas.

Let us define the density of particles with definite momentum {\bf k},
$n_k = |\beta_k(t)|^2.$ Using Eqs.~\leqn{ap:Bogol} and \leqn{ap:matrix}
and the normalization condition $|\alpha_k^j|^2-|\beta_k^j|^2=1$,
we find the change in this density during the production event at $t_j$:
\beqa
n_k^{j+1} &=& e^{-\pi\lambda} + (1+2e^{-\pi\lambda})n_k^j \CR & & 
- 2 e^{-\pi\lambda/2} \sqrt{1+e^{-\pi\lambda}} \sqrt{1+n_k^j} \sqrt{n_k^j} 
\sin \vartheta,    
\eeqa{ap:dchange}
where $\vartheta = \varphi + 2\theta_k^j + \arg \beta_k^j - \arg \alpha_k^j$. 
When the occupation numbers are large, \leqn{ap:dchange} can be 
approximated by 
\beq
n_k^{j+1} = \exp(2\pi \mu_k^j) \, n_k^j
\eeq{ap:dlarge}
where the ``growth index'' is given by
\beq
\mu_k^j = {1\over 2\pi} \ln \left( 1 + 2e^{-\pi\lambda} - 2 e^{-\pi\lambda/2} 
\sqrt{1+e^{-\pi\lambda}} \sin\vartheta \right).
\eeq{ap:growth}
This growth index characterizes the ``efficiency'' of each particle 
production event.\footnote{Note that for some values of $\vartheta$
the growth index could become negative, corresponding to a decrease of
the particle number at time $t_j$.} It was argued in \cite{Linde1} that 
in an expanding universe the phases $\vartheta$
corresponding to different production moments $t_j$ are practically 
uncorrelated, and therefore it is a reasonable approximation to treat
the phase as a random variable. The effective growth index $\mu_k$ can 
then be obtained by averaging over the phase $\vartheta$, leading to the
expression
\beq
n_k(t) \approx \frac{1}{2} \exp(2m \bar{\mu}_k t).
\eeq{ap:avgrowth}
It was demonstrated in \cite{Linde1} that the results of this approach 
agree reasonably well with more exact numerical results.

It is clear from \leqn{ap:growth} that successful particle production
requires $\pi \lambda \lapproxeq 1$. For generic initial conditions on
the inflaton, $\theta \gg \sqrt{m/M}$, this condition practically 
coincides with the condition \leqn{condSUSY} obtained from the more 
qualitative analysis of section \ref{sce0}.   
For special initial conditions, $\theta \lapproxeq 
\sqrt{m/M}$, the inflaton passes through the instability region and
the quantity $\lambda$ becomes negative at sufficiently low momenta. Since 
the growth index increases with decreasing $\lambda$, this leads to
enhanced particle production rate. (In the non-supersymmetric case, the 
maximal production rate is achieved when the bare mass of the matter scalar 
is set to zero, corresponding to 
$\lambda = \kappa^2 \geq 0$. This is the case considered in \cite{Linde1}.)
For example, if $\sin \theta =0$, the minimal value of $\lambda$ is about 
$-mM/k_*^2 \approx -M/\sqrt{g^2 S_0^2-M^2}$. If $M\ll gS_0$, the
enhancement is insignificant. For $gS_0\sim M$, however, this effect can 
become very important.  

The rates of $X_2$ particle production can be derived in the same way.     
The only difference is the sign change in the last term of \leqn{ap:lambda}.
As a result of this change, $\lambda$ is positive-definite in this case.
At any rate, as long as $M \ll gS_0$ and $m\ll M$, the produced number 
densities of $X_1$ and $X_2$ particles are nearly identical.  

Finally, let us note that the above analysis could also be performed in the
non-supersymmetric model of Section~\ref{nonSUSY}; one just has to use the
corresponding expression for the effective frequency, Eq.~\leqn{omega}, 
instead of~\leqn{ap:omega}. The results are identical to the ones 
presented here with the replacement $M^2\sin^2\theta-mM\cos\theta 
\goesto M_0^2$. (Of course this replacement does not make sense for
the special initial conditions, $\theta \lapproxeq \sqrt{m/M}$, in the 
supersymmetric model.) In particular, for generic initial conditions, 
$\theta\gg\sqrt{m/M}$, the production rates of $X_{1,2}$ bosons are the
same as for a particle of the bare mass $M\sin\theta$ in the non-supersymmetric
model. 

\subsection{$\eps\not=0$.}

Now, consider scalar preheating in the case $\eps\not=0$. As we discussed in 
Section~\ref{SCeps}, non-adiabatic particle production can only 
occur when the spiral inflaton trajectory passes through the instability 
region~\leqn{region}. Let us consider a circular trajectory satisfying this
property: $S_1=|S_*|\sin mt$, $S_2=|S_*| \cos mt$. The effective frequency
for the field $X_1$, defined in Eqs.~\leqn{ap:mode} and~\leqn{ap:omega}, is
given by
\beq
\omega_k^2 = \frac{{\bf k}^2}{a^2(t)} + 2 M^2\,(1-\sin mt) - mM,
\eeq{ap:ef1}  
where we have used $S_*=-M/g$. The inflaton passes through the center of
the instability region at times $t_j = m^{-1}(\pi/2 + j\pi).$ Since the
instability region is small, each of the particle production periods is
short compared to the inflaton oscillation period $m^{-1}$. This allows us to
Taylor expand the sine function in~\leqn{ap:ef1} for $t\approx t_j$:
\beq 
\omega_k^2 = \frac{{\bf k}^2}{a^2(t)} -mM + M^2 m^2\,(t-t_j)^2.
\eeq{ap:Taylor1}
Defining $\tau=\sqrt{Mm}(t-t_j)$, the mode equation~\leqn{ap:mode} at $t=t_j$ 
becomes identical to~\leqn{ap:modeT}, with 
\beq
\lambda_k^j = -1\,+\,\frac{{\bf k}^2}{a^2(t_j)mM}.
\eeq{ap:lambda1}
Thus, the analysis of particle production in the $\eps\not=0$ case 
effectively reduces to the analysis for $\eps=0$ performed in the previous
subsection. The change in the occupation numbers $n_k$ during a single 
production event at time $t_j$ is given by Eq.~\leqn{ap:dchange}; assuming 
large $n_k$ we obtain
\beq
n_k^{j+1} \approx (1+2e^{-\pi\lambda}-2e^{-\pi\lambda/2}\sqrt{1+
e^{-\pi\lambda}}\,\sin\vartheta) n_k^j,
\eeq{ap:dlarge1}   
where $\vartheta$ is a random phase and $\lambda\equiv\lambda_k^j$.
This formula and Eq.~\leqn{ap:lambda1} make it clear that particle 
production by an inflaton on a spiral trajectory is extremely efficient: the 
occupation numbers of states with low momenta typically grow by a factor of 
$e^{\pi} \approx 25$ in a single production event. A rough estimate of the 
energy density of the produced $X_1$ particles yields 
\beq
\rho(t)\approx \frac{m^2 M^2}{16\pi^5N^2}\, e^{\pi N}, 
\eeq{ap:rho}
where $N$ is the number of production events before $t$. This estimate suggests
that this process is not, in fact, accurately described by the stochastic 
resonance picture: after just two or three production events, the energy 
density of the $X$ particles becomes comparable to that of the inflaton, and
their backreaction becomes crucially important. Such explosive particle
production in the instability region is similar to the tachyonic
preheating~\cite{tachyonic} which occurs in hybrid inflation models.   

\section{Ellipticity of the Inflaton Trajectory.}
\label{appb}

When $\eps=0$ in the superpotential~\leqn{super}, the complex phase of the 
inflaton field $S$ is conserved, and its trajectory in the complex $S$ plane 
is a straight line. When $\eps$ is turned on, the phase is no longer 
conserved, and the trajectory is more complicated. As we have argued in 
section~\ref{SCeps}, the evolution of the inflaton can be divided into 
two stages: during the first stage the phase-violating terms are important, 
while during the second stage they can be neglected. The inflaton trajectory
during the second stage is an elliptical spiral; its ellipticity is 
determined by the amount of angular momentum accumulated in the first stage.
In this appendix we would like to argue that even for small violations of
phase invariance ($\eps \sim 10^{-6}$) the accumulated angular momentum is 
sufficient to obtain order-one ellipticity.

The inflaton potential has the form 
\beq
  V = |m S + \epsilon S^2|^2
  = m^2 |S|^2 + \eps m |S|^2 (S+S^*) + \epsilon^2|S|^4.
\eeq{ap:pot}
The Noether current for the phase invariance is 
\beq
  n_S = \int d^3 x \, i (S^* \dot{S} - \dot{S}^* S).
\eeq{ap:Noether}
Both the first and the third terms in the potential~\leqn{ap:pot} preserve 
the phase invariance, while the second term explicitly breaks it. In
the presence of the expansion of the universe, the Noether current
satisfies the equation
\beq
  \dot{n}_S + 3H n_S = i \left( \frac{\partial V}{\partial S} S
    - \frac{\partial V}{\partial S^*} S^*\right)
  = i \eps m |S|^2 (S-S^*).
\eeq{ap:Noether1}
In the case $\eps = 0$, the motion of the inflaton is described by 
\beq
  S(t) = \frac{t_0}{t} S_0 \sin mt
  = \frac{t_0}{t} |S_0| e^{i\theta} \,\sin mt ,
  \label{eq:S(t)}
\eeq{ap:line}
where $\theta$ is determined by initial conditions. (Here we assumed that 
the motion of $S$ has virialized, which is true when the amplitude is less 
than the reduced Planck scale.) For sufficiently small 
$\eps$, the violation of the phase conservation is small, and the 
inflaton trajectory is well approximated by \leqn{ap:line}. In this case, 
\leqn{ap:Noether1} can be integrated:
 \beqa
  n_S a^3 (\infty)
  &=& \int_{t_0}^\infty dt\, i \eps m |S|^2 (S-S^*) a^3(t) \CR
  &=& 2 \eps m |S_0|^3 a_0^3 t_0 \sin\theta \, 
  \int_{t_0}^\infty dt\, \frac{1}{t} \sin^3 mt,
\eeqa{ap:Noether2}
where $a$ is the scale factor. The last integral is order unity. (If the 
integration is taken from $t_0 = 0$, it is $\pi/4$.)  
 
The quantity determining the ellipticity of the inflaton orbit is
\beq
  \frac{n_S}{m |S|^2}
  \simeq 2 \eps m |S_0|^3 a_0^3 t_0 \sin\theta \, \frac{1}{m |S|^2
  a^3}
= 2 \epsilon |S_0| t_0 \sin \theta.
\eeq{ap:ellipt}
Because $t_0$ is defined to be the moment where $S$ starts oscillating
around the origin following the harmonic oscillator potential, $t_0
\simeq M_*/(m |S_0|)$ ($M_* = M_p/\sqrt{8\pi}$), we find
\beq
  \frac{n_S}{m |S|^2}
  \simeq 2 \epsilon \,\frac{M_*}{m}\,\sin\theta.
\eeq{ap:ellipt1}
The above analysis is valid only if the ellipticity is small. For generic 
initial conditions on the inflaton and $m\sim 10^{13}$ GeV, this implies 
$\eps \ll 10^{-6}$. The analysis  
breaks down for $\eps\sim 10^{-6}$, indicating that the ellipticity of
the orbit in the case studied in Section~\ref{SCeps} is of order one
unless the initial conditions are tuned so that $\theta \approx 0$.  

%%%%%%%%%%%%%%%%%%%%%%%%%%%%%%%%%%%%%%%%%%

\end{document}